\documentclass{edpsciences_aa}

\input psfig

\newcommand{\Msolar}{\mbox{\,$\rm M_{\odot}$}}        
\newcommand{\Rsolar}{\mbox{\,$\rm R_{\odot}$}}        
\newcommand{\Lsolar}{\mbox{\,$\rm L_{\odot}$}}        

\def\hp{\hphantom}
\def\pz{\hp{0}}
\def\tief #1{_{\rm #1}}
\def\hoch #1{^{\rm #1}}
\def\mc{\multicolumn}

\begin{document}


\title{High resolution spectroscopy of symbiotic stars
       \thanks{Based on observations obtained at the 
               European Southern Observatory, La Silla, Chile;
        the observations were granted for the ESO programs  
        47.7-081, 48.7-083, 49.7-041, 50.7-129, 51.7-093,
        52.7-068, 53.7-083, 54.E-061, 55.E-446, 56.E-526
}}
\subtitle{VI.~Orbital and stellar parameters for AR\,Pav}
\author{H.~Schild\inst{1} \and T.~Dumm\inst{1} \and U.~M\"urset \inst{1} 
    \and H.~Nussbaumer\inst{1} 
	\and H.M.~Schmid\inst{2}\and W.~Schmutz\inst{3} }
     \institute{Institut f\"ur Astronomie, ETH-Zentrum, CH-8092 Z\"urich, 
           	Switzerland
\and
     Landessternwarte Heidelberg-K\"onigstuhl, D-69117 Heidelberg, 
           Germany
\and                
     Physikalisch-Meteorologisches Observatorium, CH-7260 Davos,
           Switzerland}

\offprints{H. Schild, hschild@astro.phys.ethz.ch}
\date{Accepted November 28 2000}

\abstract{We present new dynamical parameters of the AR~Pav binary system. Our
observations consist of a series of high resolution optical/NIR spectra
from which we derive the radial velocity curve of the red giant as
well as its rotation velocity. Assuming co-rotation, we determine 
the stellar radius (130\,\Rsolar) of the red giant. Based on this we
derive the 
red giant's luminosity and mass (2.0\,\Msolar) as well as the distance of the 
system (4.9\,kpc). 
The binary mass function finally yields the companion's mass (0.75\,\Msolar) 
and the binary separation (1.95\,AU). We find that the red giant does not 
fill its Roche lobe.
We review the radial velocity data of Thackeray and Hutchings (1974),
and compare it with our red giant's orbit. We find that their RV curves
of the blue absorption system and the permitted emission lines are in
anti-phase with the red giant, and that the forbidden emission lines
are shifted by a quarter of a period. The blue absorptions and the permitted
emission lines are associated with the hot companion but not in a
straightforward way. The blue absorption system only tracks the hot
component's orbital motion whilst it is in front of the red giant, whereas
at other phases line blanketing by interbinary material leads to
perturbations.
We finally present UV light curves based on IUE archive spectra. We 
clearly detect eclipses in the continuum at all wavelengths. The
eclipse light curves are unusual in that they 
show a slow and gradual decline prior to eclipse which is followed
by a sharp increase after eclipse.
}

\maketitle
\titlerunning{Orbital and stellar parameters for AR Pav}
\authorrunning{H. Schild et al.}

\section{Introduction}

AR Pavonis was the first symbiotic star to be recognized as a binary system.
Mayall (1937) discovered it as an eclipsing P Cygni type star with a period of
605 days. She pointed out, how remarkable it was to observe an eclipsing star
with such a long period. She also warned that its variability was unusual
for a P Cygni star. Indeed, the association with P Cygni was only due to the
presence of Balmer lines in emission with strong absorption edges toward the
violet. As MWC 600 it entered Merrill and Burwell's (1943) catalogue of Be and Ae
stars as type Beq. Although Sahade (1949) extended the list of observed absorption
and emission lines, it was Thackeray (1959) who made a first attempt at modelling,
breaking away from the idea of a peculiar P Cygni variable. Finally, Thackeray \&\ 
Hutchings (1974) claimed AR Pav for the symbiotic family. Based on their extended 
set of spectroscopic observations, covering several eclipses, they proposed a 
binary model. The strength of the TiO bands suggested an M3 III as secondary. 
In their opinion the eclipse was not just due to the
extinction of stellar radiation, but to the ``eclipse of a region whose
excitation and density increases towards a central source of light''. They 
distinguish different radial velocity behaviours in permitted emission, forbidden 
emission, and absorption lines. Permitted emission lines, e.g. \
HeII $\lambda4686$, give the clearest eclipse pattern. They associate it with nebular
emission centered on what to them is the hot primary, and they derive a nearly
circular orbit and a mass function of $m\tief f\hoch {hot}$=0.14. Based on
considerations about the eclipse length and the probable mass of the secondary,
they estimate a mass ratio of $q\sim$~2 and suggest 
a hot ($T\approx30\,000$\,K) star with a mass of $\approx2$\,\Msolar, or more precisely
$M\tief{hot}= 2.5$\,\Msolar, $M\tief{cool}= 1.2$\,\Msolar. 

The same mass function 
was adopted by Skopal et al.~(2000) when they analysed the 1889-1998 light
curve to arrive at $M\tief{hot}= 4.5$\,\Msolar, $M\tief{cool}= 1.8$\,\Msolar. Thackeray \&\
Hutchings (1974) do find less clear velocity patterns in the forbidden emission or 
the absorption lines in the blue continuum. They associate them with gas streams 
(see Fig.\,5 in Thackeray 
\& Hutchings), and do not derive a velocity curve for the cool giant.
In this paper we question the assumption of Thackeray \& Hutchings (1974) that the
allowed emission lines accurately trace the orbit of the hot star. 

\begin{figure}
\hbox{\psfig{figure=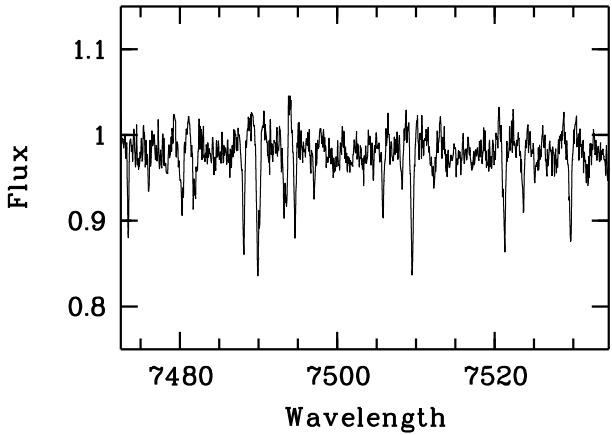,height=6.0cm,width=8.5cm,clip=}}
\hskip-0.5cm
\caption[]{Example of a normalised red spectrum of AR~Pav that served
for cross correlation with an RV standard star.}
\label{spec.ima}
\end{figure}

We have observed a set of absorption lines from the red giant (Fig.~\ref{spec.ima}) over 
3.5 binary periods. From this we
derive an orbital solution, including a revised mass function. Our observations
also permit us to extract the rotation velocity of the red giant, and with the
assumption of corotation we derive the absolute radius of the red giant, its 
luminosity, and the distance to AR Pav. 

The analysis of Thackeray \& Hutchings (1974) as well as our analysis are
based on the assumption that the occurence of an eclipse implies an approximate
coincidence of the line of sight with the binary orbital plane of AR Pav.
However, this has been contested by Hutchings et al. (1983). They claim that 
IUE (International Ultraviolet Explorer) observations prove that the hot star is
not eclipsed. We re-anaylize the IUE data and establish that eclipses are
clearly present.

Our orbital analysis of the red giant spectrum follows the same pattern as was 
already applied to the symbiotic systems 
SY~Mus (Schmutz et al.\ 1994, Paper~I),
RW~Hya (Schild et al.\ 1996, Paper~II),
CD--43.14304 (Schmid et al.\ 1998, Paper~III),
BX~Mon (Dumm et al.\ 1998, Paper~IV), and FG~Ser (M\"urset et al. 2000, Paper~V).

\begin{table}
\caption[]{Measured radial velocities and log of observations for AR~Pav. 
The phase $\phi$ is
calculated from the orbit solution given in Table~\ref{bahnparam}. RV is
the measured heliocentric radial velocity of the cool giant. O--C are the
residuals in [km/s]. The Julian
date  is relative to 2\,400\,000.}
\label{obslog}
\begin{center}
\begin{tabular}{rrrrcccr}
\hline\noalign{\smallskip}
\multicolumn{3}{c}{Date} & JD~~ & $\lambda\tief c$ & $\phi$ & RV & O--C\\
d & m &	yr&& [\AA]     & &[km/s] &\\
\noalign{\smallskip}\hline\hline\noalign{\smallskip}        
 12& 9&91 &  48511.6 & 7005 & 0.62 & $ -75.5 \pm 1.0$ &--0.7\\ 
 24&10&   &  48553.8 & 6565 & 0.69 & $ -77.4 \pm 1.0$ &--0.2\\ 
 15& 5&92 &  48757.7 & 6565 & 0.02 & $ -68.3 \pm 1.0$ &--1.3\\ 
 24& 5&93 &  49131.9 & 6565 & 0.64 & $ -77.9 \pm 1.0$ &--2.0\\ 
 29&10&   &  49289.5 & 6565 & 0.90 & $ -74.7 \pm 1.0$ &--0.8\\ 
 31&10&   &  49291.5 & 6825 & 0.91 & $ -73.6 \pm 1.0$ &0.1\\ 
  1&11&   &  49292.5 & 5880 & 0.91 & $ -75.0 \pm 1.0$ &--1.4\\ 
 17& 7&94 &  49550.6 & 6565 & 0.34 & $ -61.4 \pm 1.0$ &--1.3\\ 
 12& 8&   &  49576.6 & 7505 & 0.38 & $ -60.6 \pm 1.5$ &1.1\\ 
 13& 8&   &  49577.7 & 8170 & 0.38 & $ -62.7 \pm 1.5$ &--0.9\\ 
  1&11&   &  49657.6 & 7005 & 0.51 & $ -70.4 \pm 1.0$ &--1.2\\ 
 24& 5&95 &  49861.8 & 6565 & 0.85 & $ -77.4 \pm 0.9$ &--1.2\\ 
 29& 5&   &  49866.8 & 7453 & 0.86 & $ -74.7 \pm 0.5$ &1.2\\ 
 16&11&   &  50037.5 & 7005 & 0.14 & $ -61.8 \pm 1.0$ &--0.8\\ 
 17&11&   &  50038.5 & 6825 & 0.14 & $ -60.7 \pm 0.7$ &0.2\\ 
 31& 3&96 &  50173.9 & 6565 & 0.37 & $ -58.2 \pm 1.5$ &3.0\\ 
  2& 7&   &  50266.7 & 5007 & 0.52 & $ -70.0 \pm 0.6$ &--0.4\\ 
 22& 7&   &  50286.6 & 5007 & 0.55 & $ -71.6 \pm 0.7$ &--0.1\\ 
  7& 8&   &  50302.7 & 5007 & 0.58 & $ -71.0 \pm 0.6$ &2.0\\ 
  7& 8&   &  50302.8 & 4363 & 0.58 & $ -73.6 \pm 1.4$ &--0.6\\ 
  8& 8&   &  50303.7 & 7005 & 0.58 & $ -73.6 \pm 0.6$ &--0.5\\ 
 20& 7&97 &  50649.8 & 5007 & 0.15 & $ -57.8 \pm 1.5$ &2.7\\ 
\noalign{\smallskip}
\hline
\noalign{\smallskip}
\end{tabular}
\end{center}
\end{table}

\section{Observations}

During an observational campaign dedicated to symbiotic stars
we collected high resolution spectroscopic data of AR~Pav over
a period of six years. The observations
were carried out with the 
1.4\,m Coud\'e Auxiliary Telescope (CAT) at ESO's
La Silla Observatory in Chile. The telescope fed a Coud\'e Echelle 
spectrograph which was set to provide a spectral resolution
$R$ of 60\,000 -- 100\,000. The spectral coverage was 
approximately 60~\AA. The data were recorded with CCD
detectors that changed as time went by (ESO\#9, ESO\#30, ESO\#34,
ESO\#38). Most observations were carried out remotely from the ESO 
headquarters at Garching near Munich. A log of our observations is given in 
Table~\ref{obslog}. We observed at different wavelengths.
The red/NIR settings at 6825\,\AA, 7005\,\AA, 7505\,\AA, and 8170\,\AA\ were
devised to record the red giant absorption spectrum, whereas the
optical settings at 4363\,\AA, 5007\,\AA, and 6565\,\AA\ were meant to record 
nebular emission line profiles. In all
settings the absorption spectrum of the M giant was clearly present
and could be used to get radial velocity data.
The nebular emission lines are much broader than the M star absorption
features, and were
removed with a low-pass filter before cross-correlation.

\begin{figure*}
\hskip-0.2cm
\hbox{\psfig{figure=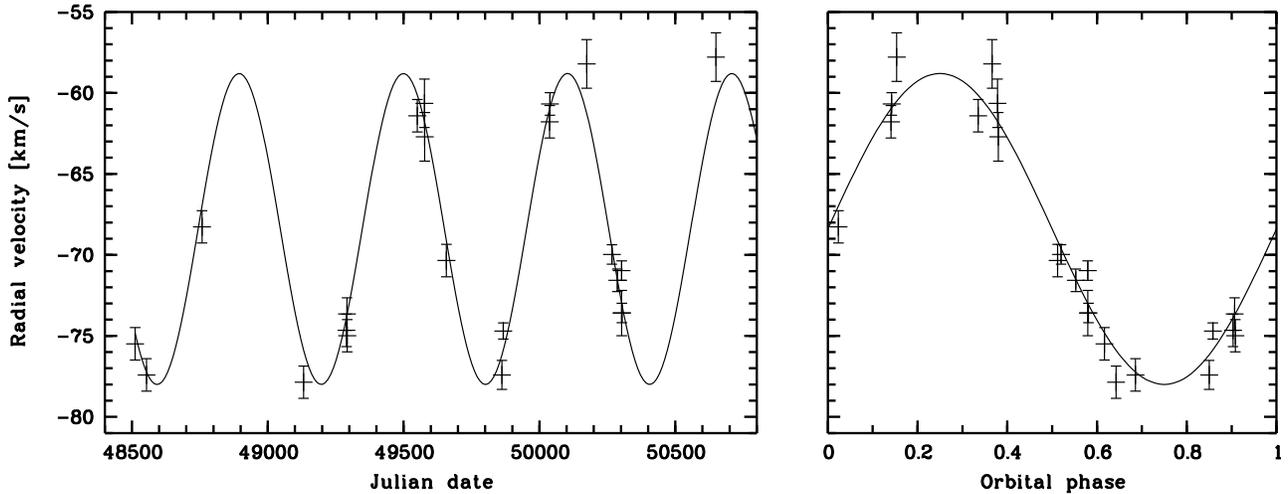,height=7.3cm,width=18.0cm,clip=}}
\caption[]{Heliocentric radial velocity of the red giant in AR~Pav with
the orbital solution from Table 2.}
\label{orb.ima}
\end{figure*}

The observations were reduced in a standard way with the Midas 
software package. Particular attention was paid to the wavelength
calibration, because this crucially influences radial velocity measurements.
As an example of the resulting data we show in Fig.~\ref{spec.ima} 
the normalised spectrum obtained in the night of August 12, 1994.

We also observed the radial velocity standards BS\,4763,
BS\,6973 and BS\,8795 with the same 
instrumentation. Their radial velocities were taken from the
Astronomical Almanac for BS~4763, and from Beaver \& Eitter (1986),
Barnes et al. (1986) and Beavers et al (1979) for the other standard stars.
Cross-correlating
the spectra of AR Pav with standard stars usually yielded
robust correlation peaks with a FWHM of approximately 0.2\,\AA. Fitting a Gaussian
profile to these peaks allowed the center to be determined to an accuracy
of about 0.02\,\AA. This transforms to a radial velocity accuracy of 
$\approx$\,1\,km/sec. In cases where the correlation peak is much broader or 
weaker than
normal, we adopted a larger error of 1.5\,km/s. Occasionally we obtained more
than one spectrum per night. Combining them yielded
RV values with smaller errors (Table~\ref{obslog}).

\section{The cool giant}
\subsection {Orbit}

The period of AR~Pav of 604.5\,d is well established by 
photometry dating back to the year 1889 (Mayall 1937).
A time series analysis of our radial
velocity data yields a period of 605\,days. We adopt the more precise photometric 
value of Bruch et al. (1994), and force our orbital solution to a period of 
604.5\,d. 
A least squares fit with the remaining orbital parameters left free, leads 
to a best solution with a slight eccentricity of
$e = 0.04$. The (O-C) residuals for a circular orbit are, however,
not significantly larger and, given the fact that a circular orbit requires 
two fitting parameters less, it provides an even better 
description of the data. A circular orbit is also
in agreement with the tidal theory
of Zahn (1977) which predicts a circularisation time of less than 50\,000~y. 
Our adopted orbital solution is plotted in Fig.~\ref{orb.ima}, and the
corresponding
orbital parameters are listed in Tab.~\ref{bahnparam}. The residuals
O--C of the fit are given in Tab.~\ref{obslog}.

The fact that the best orbit is close to circular is very satisfactory
not only because it is expected theoretically, but also because it
demonstrates that our RV measurements are not affected by a phase dependent
mass outflow. Such an outflow may well be present in the AR Pav system
(see Sect. 5.2), but apparently it is not capable of modifying the spectrum 
of the red giant because its effect would be to mimic an
eccentric orbit. We also
note that the cross correlation technique is insensitive to small
spectral changes as e.g. irradiation from the companion might produce.

With our new orbit the ephemeris for the lower conjunction of the red 
giant ($\phi$ = 0) is 

\begin{equation}
2448139 + 604.5 \times n \ .
\end{equation}
This conjunction phase closely matches the photometric minima
(see e.g. Skopal et al. 2000) which proves that the red giant is the
eclipsing body.

From the orbital parameters of the cool giant we compute the mass function
\begin{equation}
   m\tief f \hoch{cool} = {1\over2\pi G}PK\hoch 3 \left(1-e\hoch 2\right)\hoch {3/2} 
	= 0.055\,\Msolar\ .
\end{equation}
$P$ is the orbital period, $K$ the velocity semi-amplitude, $e$ the eccentricity, 
and $G$ the gravitational constant. This mass function is linked via

\begin{equation}
m\tief f \hoch{cool} = {\left(M\tief h\cdot\sin i\right)\hoch 3\over\left(M\tief h+M\tief c\right)\hoch 2} 
\end{equation}
to $M\tief h$ and $M\tief c$, the masses of the hot and the cool
components, respectively. As the system is eclipsing, the orbit inclination $i$ 
is certainly high. From the UV eclipses we estimate that 
$i>80\hoch {\circ}$ (see Sect.~\ref{uvlight}).

\begin{table}[h]
\def\ul{\underline}
\caption[]{Orbital solution for the cool giant in AR~Pav. 
Underlined values are pre-determined (see text). $P$: binary period, $T\tief0$: 
date when the red giant is in front of the hot component ($\phi=0$); 
$K$: radial velocity semi--amplitude; $V\tief0$: system radial velocity.
}
\label{bahnparam}
\begin{center}
\begin{tabular}{lcc}
\hline\noalign{\smallskip}
\ts\ts  &\mc2c{orbital solution}\\
\noalign{\smallskip}\hline\hline\noalign{\smallskip}
\ts\ts $P$ [d]                		&$\ul{604.5}$	\\
\ts\ts $T\tief 0$ [JD]             		&48139$\pm 4$ \\
\ts\ts $K$ [km\ts s$\hoch {-1}$]   		&9.6$\pm 0.4$ \\
\ts\ts$V\tief 0$ [km\ts s$\hoch {-1}$]  		&--68.4$\pm 0.3$ \\
\ts\ts $e$         			&$\ul{0}$\\
\noalign{\smallskip}
\ts\ts $m\tief f \hoch{cool}$~[\Msolar] &0.055$\pm 0.007$ \\\\
\noalign{\smallskip}
\hline
\end{tabular}
\smallskip
\vskip-0.5cm
\end{center}
\end{table}

\subsection{Radius}

As outlined in previous papers of this series, it is possible to measure
the rotation velocity of the red giant by a line profile fitting procedure 
(see Paper I). In the case of AR Pav we find a rotation velocity 

$$
v\tief{rot}\hoch {proj} = v\tief{rot}~{\rm sin}\,i = 11\pm2~{\rm km/s} \ .
$$

Since AR Pav is an eclipsing binary system, the projected rotation
velocity is close to the real rotation velocity. The error due to
the inclination uncertainty is at most a few percent and negligible
compared to the measurement error in $v\tief{rot}\hoch {proj}$. We now further assume
that the system is co-rotating. In this case the absolute radius 
immediately follows from

\begin{equation} 
R\tief c=\frac{P\cdot v\tief{rot}}{2\pi}=130\pm25~\Rsolar
\;.\end{equation}

The assumption of co-rotation is theoretically well supported. According
to the tidal theory of Zahn (1977), the synchronization time scale for a binary
system like AR\,Pav is very short, and of the order of 10$\hoch 4$\,y. 

We note that our measured radius
of the red giant is close to the median radius of 120\,\Rsolar\ 
of giants of spectral type M5 (Dumm \& Schild 1998).

\subsection{Luminosity, mass and distance \label{redmass}}

According to M\"urset \& Schmid (1999) the red giant in AR Pav is of 
spectral type M5. This corresponds to an effective temperature $T\tief{eff}$ = 
3470\,K (Dyck et al. 1996) which, together with our stellar radius
from above, yields a luminosity of 2200$\,\pm$\,1100 \Lsolar. 
We note
that this luminosity does not depend on the distance because it is based
on the stellar radius from the measured rotation velocity
(assuming co-rotation). The relatively large luminosity error 
is almost entirely due to the uncertainty in the red
giant radius $R\tief c$. Plotting
the red giant stellar parameters in a HR diagram together with RGB/AGB 
evolutionary tracks from Bressan et al. (1993) yields for the red giant's
mass $M\tief c$ = 2.0$\,\pm$\,0.5\,\Msolar. Due to the fact that the evolutionary
tracks are very steep, the large error in the
red giants' luminosity translates into an acceptable uncertainty in
the stellar mass.

The known radius of the red giant opens the way for a
distance determination. From the surface brightness
relation for M-giants given in Schild et al.\ (1999) we
derive a distance

\begin{equation}
d = R\tief c\cdot10\hoch {7.96+0.33\cdot{\rm K}-0.13\cdot{\rm J}} = 4.9\,{\rm kpc} 
\end{equation}
by inserting the mean observed IR magnitudes from Glass \& Webster
(1973). The error in the distance (Tab.~\ref{results}) is again dominated
by the uncertainty of the red giant radius $R\tief c$.
The above distance  is in good agreement
with previous estimates which range from 3.8\,kpc to 5.8\,kpc (Thackeray \&
Hutchings 1974, Kenyon \& Webbink 1984, Skopal et al. 2000).

\begin{table}
\caption{Summary of the parameters derived for the AR~Pav system.}
\label{results}
\begin{center}
\begin{tabular}{lrr}
\hline\noalign{\smallskip}
Parameter					&Adopted 	&\mc1c{Uncertainty}	\cr
\noalign{\smallskip}\hline\hline\noalign{\medskip}
\underline{System parameters:}\cr
\noalign{\smallskip}
\ts\ts Distance $d$ [kpc]			&4.9\pz\pz	&1.2\pz\pz\pz		
\cr
\ts\ts Period $P$ [d]				&604.5$\hoch *$\pz	&		
\cr
\ts\ts Eccentricity $e$				&0\hp{.000}	&$<0.06$\pz\pz		
\cr
\ts\ts Separation $a$ [AU]			&1.95\pz	&0.15\pz\pz		
\cr
\ts\ts Mass function $m\tief f \hoch{cool}$ [\Msolar]	&0.055	&0.007\pz		
\cr
\ts\ts Total mass [\Msolar]			&2.75\pz	&0.65\pz\pz
\cr
\ts\ts Mass ratio                               &2.7\pz\pz      &0.4\pz\pz\pz
\cr
\noalign{\medskip}
\underline{Cool component:}\cr
\noalign{\smallskip}
\ts\ts Mass $M\tief c$ [\Msolar]		&2.0\pz\pz	&0.5\pz\pz\pz		
\cr
\ts\ts Radius $R\tief c$ [\Rsolar]		&130\hp{.000}	&25\hp{.0000}		
\cr
\ts\ts Luminosity $L\tief c$ [\Lsolar]		&2200\hp{.000}	&1100\hp{.0000}		
\cr
\noalign{\medskip}
\underline{Hot component:}\cr
\noalign{\smallskip}
\ts\ts Mass $M\tief h$ [\Msolar]		&0.75\pz	&0.15\pz\pz		
\cr
\noalign{\smallskip}\hline
$\hoch *$ adopted
\end{tabular}
\end{center}
\end{table}

\section{The hot component and the binary geometry}

\subsection{Mass of the hot component}

The mass function of Tab.~\ref{bahnparam} together with the red giant's mass from
Sect.\,\ref{redmass} yield a mass of the hot component of 
$M\tief h$ = 0.75\,\Msolar. This is slightly higher than the canonical mass of a single 
white dwarf, but still close to that value. Thus, also in the AR~Pav symbiotic system 
the companion of the red giant is most probably a white dwarf. However, it is noteworthy 
that the hot
companion in AR~Pav is more massive than in any other previously studied symbiotic 
system. On average the hot companions in symbiotic systems tend to have masses 
lower than the canonical WD mass, at around 0.50\,\Msolar\ 
(e.g. M\"urset et al. 2000).

\subsection{Binary configuration}

With both stellar masses known, we can now completely establish the binary 
configuration. Kepler's third law yields a binary separation of 1.95\,AU. 
The distance between the center of the red giant and the inner Lagrange point is 1.2\,AU. 
The red giant extends to
approximately 50\% of the Roche radius. Among the symbiotics with well established 
orbits, AR~Pav comes
closest to filling its Roche lobe. Although mass transfer via Roche lobe
overflow is unlikely, it is expected that the hot companion 
accretes a fraction
of the red giant wind. We will come back to the consequences
of wind accretion in Sec.~\ref{discussion}.

For a point--like hot component the eclipse duration in the case of  
central passage would last 60\,$\pm\,20$\,d. This is in reasonable
agreement with the observed mean eclipse duration of 82\,d by
Bruch et al. (1994).

\section{Earlier radial velocity curves}

We now compare our radial velocity curve of the red giant with earlier RV measurements 
of Thackeray \& Hutchings (1974), who collected high resolution spectroscopic data between 
1953 and 1973. Their data set consists of the radial velocities of permitted and forbidden 
emission lines (PE and FE), and of a system of absorption lines in the
blue continuum (A). The spectra were recorded on photographic plates and covered the 
blue spectral range.

\begin{figure}
\vskip-0.2cm
\hskip-0.3cm
\mbox{
	\mbox{\psfig{figure=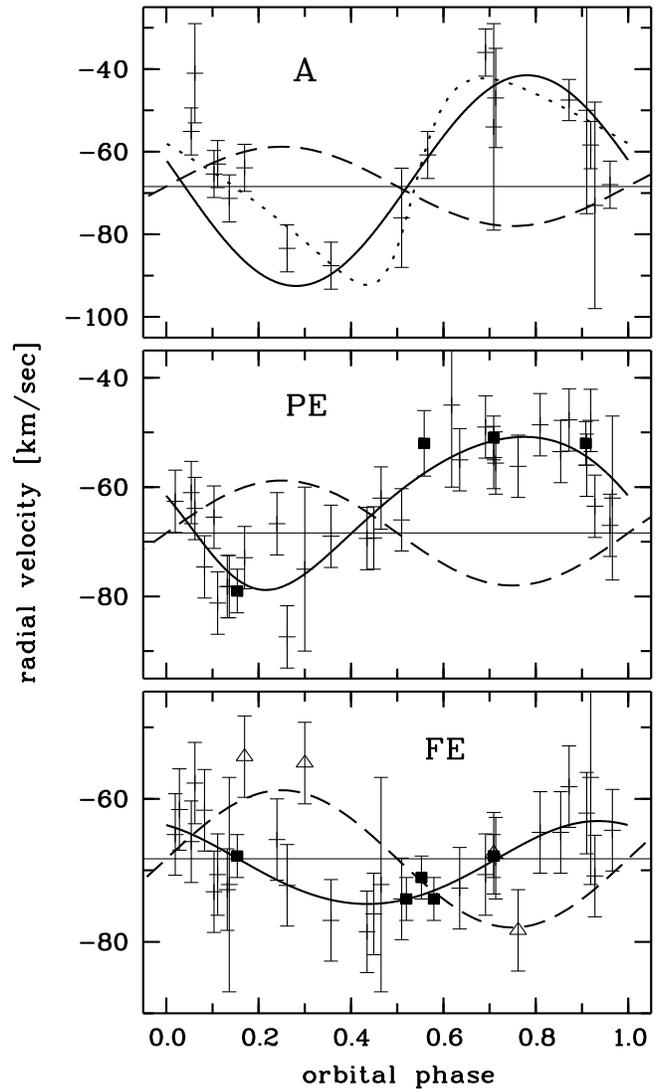,height=15cm,width=9cm,clip=}}}
\caption{Radial velocities of various spectral features
(blue absorptions, permitted and forbidden emission lines)
compared to our red giant orbit (dashed curve). 
The data of Thackeray \& Hutchings (1974) are shown as horizontal bars
or triangles. The horizontal line
represents the system velocity. 
{\it Top:} The blue absorption system A: The dotted line shows the 
best elliptical orbital fit to all data points and the
full line gives a fit restricted to the points between phase 
0.25 and 0.75 (hot component in
front). {\it Center:} Radial velocities
of permitted emission lines PE with best orbital fit. 
The filled squares show 
radial velocities from our data and from
Van Winckel et al. (1993).
{\it Bottom:}
forbidden emission lines FE. Triangles
were excluded from the fit (see text) which is plotted as a solid line. 
Filled symbols as in
central panel.
}
\label{thack.ima}
\end{figure}

\subsection{Absorption lines}

AR Pav went into an outburst in 1954. Thackeray \& Hutchings (1974) noted
that the absorption
system then became unusually strong. Subsequently the features weakened but
remained visible throughout the observation cycle. The spectra covered
the blue spectral range where the red giant was extremely weak
and the outbursting companion very bright. It was already clear to
Thackeray \& Hutchings (1974) that this blue
absorption system was linked to the hot companion. Comparison
with our red giant radial velocity curve shows indeed that
it is in anti-phase with the red giant 
(Fig.~\ref{thack.ima}, top). 

A similar pattern of two absorption systems in anti--phase
was previously seen during outbursts of the symbiotic systems
AX Per and BX Mon (Mikolajewska \&
Kenyon 1992, Dumm et al. 1998). One absorption system, visible in the
blue was typical of an
A or F star and the other system, in the red, resembled an M type absorption
spectrum. The blue system was believed to be associated with the hot component and
the red system with the M giant.
In such a situation the mass ratio of the binary
components can be directly measured. With the inclination $i$ known, also
the individual masses can be inferred.

In the case of AR Pav, however, complications occur because the
orbital solution for the Thackeray \& Hutchings A system leads to an eccentricity
of 0.45 (with a large error) and a system velocity of --63.2\,km/sec.
Our red giant orbit, on the other hand, requires that the hot companion's
orbit is circular. We have re-calculated an orbital solution to
Thackeray \& Hutchings' data
and found $e(A) = 0.42$
and $V\tief 0(A)=-64$\,km/s, largely in agreement with the earlier result.
We conclude that the blue absorption spectra are not a straightforward reflection 
of the orbit of the hot companion.  
In Fig. \ref{thack.ima} we see that all data points that
force an elliptical orbit lie close to the eclipse of the hot companion 
(around phase 0).  We also note that $\omega$, the periastron angle, lies close to 
the line of sight, and the semi-major axis apparently is pointing towards the observer. 
These are indications that the apparent ellipticity  may be an effect tied to our line 
of sight. If the center of the hot companion lies further from us than the center of 
the red giant, absorption features intrinsic to the hot companion could be modified 
as they propagate through the dense wind of the red giant. This affects the distant 
section of the orbit more than the one closer to us. Line of sight effects can thus 
be mistaken for signatures of the orbital motion.
 
In order to exclude this possibility we can employ only those observations
where the hot source is closer to us than the red giant. From these
observations a circular orbit results that
is almost exactly in antiphase ($\Delta\phi$=0.53) with the red giant,
and which also reproduces the system velocity of the red giant curve
(Fig. \ref{thack.ima}, top, full line). 
We therefore believe that
the best interpretation of Thackeray's A system is, that it indeed
reflects
the hot companion's motion. But whilst the light from hot source passes through 
the red giant's wind,
the blue absorption features are contaminated by additional absorptions 
due to the red giant wind or a flow of material towards the hot companion.
Similar line blanketing of the hot component continuum was recently seen
in RW Hya (Dumm et al. 1999). In the case of AR Pav, 
these additional absorptions affect the radial velocity curve in such a way
that an elliptical orbit is mimicked.

\begin{figure}
\mbox{\hskip-0.2cm{\psfig{figure=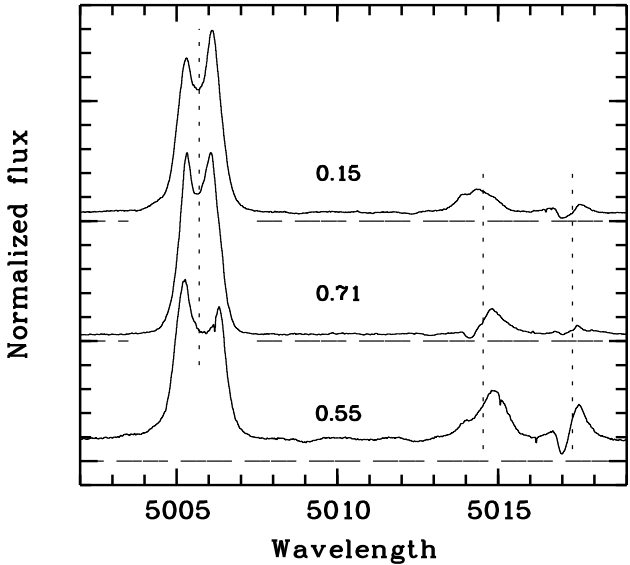,height=8cm,width=9cm,clip=}}}
\caption{The [O\,{\sc iii}]$\lambda$5007, 
He\,{\sc i}$\lambda$5015 and the Fe\,{\sc ii}(42)$\lambda$5018 emission lines
at different orbital phases 0.15 (top), 0.71 (center) and 0.55 (bottom). 
The vertical dashed lines give the zero
velocity wavelength in the AR Pav system. The central
spectrum is from Van Winckel et al. (1993) and the other two are from our data set.
}
\label{oiii.ima}
\end{figure}

With this hypothesis, it is straighforward to deduce the hot star's mass function.
The velocity semi-amplitude of the restricted circular orbit $K\hoch {hot}$ = 25.5\,km/s 
is almost unchanged with respect to Thackeray \& Hutchings original elliptical 
solution. This is not
surprising since in quadrature the orbital motion is at its highest and
absorption effects are unlikely to dominate. 
The mass function with this K value and zero
eccentricity is

\begin{equation}
m\tief f \hoch {hot} = 1.1~\Msolar \ .
\end{equation}

This value is substantially higher than the one of Thackeray \&
Hutchings (1974) which, however, was based on emission line radial velocities. 

Combining the velocity semi-amplitudes of the hot and cool stars yields a mass
ratio of $q\,=\,$2.65 which, with the assumption of sin\,$i$ = 1, leads to
individual masses $M\tief h$ = 0.75\,\Msolar\ and $M\tief c$ = 1.9\,\Msolar.
The very close agreement with our values of 
Tab.~\ref{results} is a strong and independent confirmation
of our previous method to establish binary masses in symbiotic systems.

\subsection{Emission lines}

Thackeray \& Hutchings (1974) also gave a list of radial velocities
of allowed (mainly He\,{\sc i}) and forbidden emission
lines. To these we added measurements from our observations as well as the
data published by Van Winckel et al. (1993). As an example of these 
newer data we plot in Fig.~\ref{oiii.ima} spectra which show 
the [O\,{\sc iii}]$\lambda$5007, 
He\,{\sc i}$\lambda$5015 and Fe\,{\sc ii}(42)$\lambda$5018 lines.

For the permitted lines 'PE',
an orbit with a small and insignificant ellipticity results, 
which is in clear anticorrelation with the
red giant curve (Fig.~\ref{thack.ima}, center). 
This shows that the dominant source of
permitted line emission is associated with the hot component. The
semi amplitude of 14 km/sec is however only about half of the value of
the hot component, and $V\tief 0(PE)$ of --63 km/sec is offset by about +5 km/s
with respect to the system velocity from the red giant curve. 
This offset can easily be explained by
P Cygni like absorptions which shift the line center
towards the red (see Fig.~\ref{oiii.ima}). Also the He\,{\sc i}$ \lambda$4471
profile in Thackeray \& Hutchings (1974)
shows such a structure. The P Cyg absorption does not necessarily need 
to stay constant over an orbital cycle
(Fig.~\ref{oiii.ima}) which may explain
the discrepant K value.
In addition there could be more than one PE line region
and contamination from these may affect both,
the K value as well as the system velocity.
In particular, if there was a contribution from
material in the neighbourhood of the red giant, the semi-amplitude
would be reduced. In view of these uncertainties it is highly
unlikely that the PE-spectra
directly reveal the orbit of the hot component.

The forbidden lines 'FE' show a small, but still significant periodic 
RV variation and the
mean velocity $V\tief 0(FE)$ = --69.2 km/s is very close to our red giant system 
velocity. This is expected because forbidden lines do not suffer 
self--absorption. 
The [O{\sc iii}]$\lambda$5007 lines in our data set are double--peaked with 
a peak separation of about 1\,\AA\ (Fig.~\ref{oiii.ima}) wheras 
[O{\sc iii}]$\lambda$4363
in Thackeray \& Hutchings (1974) is single--peaked. This difference in
line profile is
probably not real because the dispersion of the earlier data was not high enough
to resolve the double--peak structure.
Interestingly, the lines do not shift as a function
of phase but the relative size of the blue and red peaks vary
such that the line center appears to move backwards and forwards.
At present we do not have sufficient data to disentangle the true motion
of the various line components. We however note that the
FE radial velocity curve is not in anti-phase with the red giant orbit
but shifted by about $\Delta\phi$=0.25.
The radial velocity maxima therefore occur close to occultation in
such a way that, if the red giant is in front, the FE emission is redshifted
at maximum.

We finally note that three data points taken between JD 34942.4 and
JD 35300.4 strongly deviate from the radial velocity fit and in fact are
consistent with
the red giant orbital motion. These observations were taken during the
height of the 1954 outburst, and we can interpret the deviations in
the sense that the FE region was then much closer to the red giant.
Thackeray and Hutchings (1974) were unable to find a meaningful orbital
solution for their full set of FE data points, the iteration led to the
absurd result of e $>$ 1. If we remove these three observations we find a solution 
with an unsignificant ellipticity of
e $\approx$ 0.05 and a semiamplitude of 5.8 km/sec. The phase 
shift relative to the red giant's orbit remains unchanged at a quarter
of a period.

\section{UV light curves from IUE spectra \label{uvlight}}

We have analysed all IUE (International Ultraviolet Explorer) spectra 
of AR Pav stored in the final IUE archive. The spectra were obtained 
from 1978 to 1984 except for two observations taken in 1992. 
These last two spectra were excluded from our analysis in order 
to prevent a distortion in the eclipse curves due to the long term 
variability of AR Pav. Indeed the flux levels from the latter two
spectra do not fit the 
eclipse curve defined by the other UV data.
The few existing small aperture spectra were also disregarded because 
more accurate large aperture spectra taken on the same dates
were available.

\begin{figure}
\vskip-0.2cm
\mbox{\hskip-0.2cm{\psfig{figure=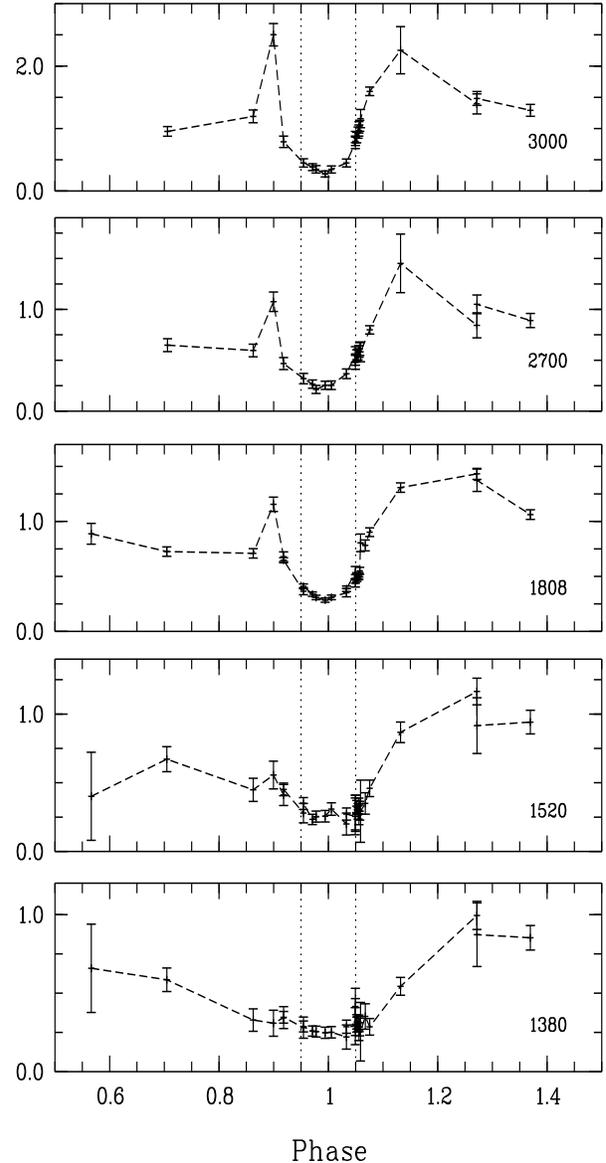,height=16cm,width=9cm,clip=}}}
\caption{UV continuum flux variations from IUE spectra. The flux
scale is in units of $10\hoch {-13}$ erg cm$\hoch {-2}$s$\hoch {-1}$\AA$\hoch {-1}$.
The dotted 
lines mark the beginning and end of the geometrical eclipse.  
}
\label{uvcont}
\end{figure}

In our search for a wavelength dependent eclipse behaviour  we measured 
the continuum fluxes at five different wavelengths (Fig.~\ref{uvcont}). 
All five light curves show a significant flux reduction around $\phi=1$. 
However, the continuum is not completely eclipsed. 
At mid-eclipse, $\phi=1.0$, there remains a nearly wavelength independent 
flux of approximately $2.5\times 10\hoch {-14}$ erg cm$\hoch {-2}$s$\hoch {-1}$\AA$\hoch {-1}$.

At all wavelengths the entry into eclipse is less abrupt then the 
recovery which coincides well with the end of the geometrical eclipse at 
$\phi=0.05$. An 
asymmetric evolution of the continuum flux through an eclipse has 
been observed in other symbiotic systems (e.g. Dumm et al. 1999)
but in AR Pav the asymmetry is more pronounced then in others.

With our orbital parameters the exit from eclipse could be 
reproduced by a pointlike UV-source on a $i= 90\hoch {\circ}$ orbit around a 
red giant of $R\tief c$= 130\,\Rsolar. This agrees well with the red giant 
radius derived from corotation (Tab.~\ref{results}). From the corresponding upper 
limit of corotation, 155\,\Rsolar, the observed exit from eclipse of 
a pointlike source would require an inclination of $i>79\hoch {\circ}$, 
or sin\,$i> 0.98$.

Contrary to other S-type systems like EG And, RW Hya, or SY Mus, the UV-continuum 
of AR Pav does not resemble that of a hot white dwarf. Out of eclipse it 
is strongest at the red end of the IUE range. This suggests a hot source 
which is considerably cooler than found in other symbiotics. 

Hutchings et al. (1983) analysed the 1982 large aperture IUE spectra. 
Their conclusions were ``. . . there is no total eclipse for less than 
$\lambda3600$; 70\%\ of the hottest continuum (below 1400 \AA) is seen 
even at central eclipse. Thus it seems that the source of hottest radiation 
(and thus presumably the central star) is not eclipsed.'' The sample of 
spectra in Hutchings et al. (1983) is narrowly spaced around mid-eclipse, 
except for an observation at $\phi=0.915$. In our Fig.~\ref{uvcont} 
we see that because of the asymmetry of the eclipse, a measurement at $\phi=0.915$ 
does not represent the uneclipsed spectrum. The restricted set of their 
observations, away from the center of the eclipse, is therefore an obvious 
source of our different conclusions. 

The present investigation concentrates on the geometrical configuration 
and the orbital parameters of the binary, but not on the nebular or 
stellar physics of the system. 
The wavelength--dependent occultation behaviour contains information about
the structure of the hot component which, however, to explore is beyond
the scope of this paper. Here we just aim to establish that AR~Pav is
indeed an eclipsing system and that our previously made assumptions about
a high orbital inclination were justified.
As the observations in the visual 
domain lead us to adopt an orbital plane with $i\approx90\hoch {\circ}$, 
that model also has to be compatible with IUE observations, in particular 
with the expected shape and depth of the eclipse. The shape of the 
eclipse shown in Fig.~\ref{uvcont} is compatible with an eclipse. 
The continuum 
is not completely eclipsed, however, this could be due to a nebular 
recombination continuum which is emitted from a volume much larger than 
that occupied by the eclipsed hot star.  The fact that the remaining
uneclipsed continuum is almost wavelength-independent is a strong
indication that it is indeed due to nebular recombination. In the
case of a partial eclipse of a disk structure such a pattern is not 
easily accomodated.

\section{Discussion\label{discussion}}

We present in this paper the first direct radial velocity measurements
of the red giant in the AR~Pav system. The red giant's orbit is close to circular
and yields a mass function of 0.055\,\Msolar. We also measure the red
giant's rotation velocity and, assuming co-rotation, deduce its radius
from which the luminosity and distance immediately follow. Based on 
RGB/AGB stellar 
evolutionary tracks we obtain the red giant's mass, and from the above
mass function the mass of the hot companion. It turns out that the AR Pav stellar
components have masses similar to other symbiotic systems. In particular
the mass of the companion is close to the canonical value of
single white dwarfs. The red giant does not fill its Roche lobe.

The velocity amplitude of the hot component can be obtained from historically 
recorded radial velocities of absorption lines in the blue continuum, 
taken during an 
outburst. Combining that result with our new radial velocities of the red 
giant yields directly the masses of the binary components. These masses are
in very close agreement with those that we derive with our
co-rotation method developed and applied in Papers I to V, as well as in this
work (Section 3 and 4). We therefore have now the first
independent confirmation that for symbiotics, this method leads to
valid results.

Our masses for the AR~Pav binary components differ strongly from
earlier values found in the literature. Thackeray and Hutchings (1974) favoured a 
value of approximately 2.5\,\Msolar\ for the
hot companion and 1.2\,\Msolar\ for the red giant. They
assumed that the radial velocity curve of the permitted emission
lines reflected the motion of the hot companion. Although this
emission is certainly associated with that object, it is not 
clear how exactly they are linked. The interpretation of emission lines involves 
a number of complicating factors like self-absorption which 
may be phase
dependent, and unknown nebular motions which may overlay the binary
orbit. In addition, different nebular emission regions might contribute with
variable fractions to the total intensity. We conclude that in symbiotic systems, 
emission lines are rather poor tracers of stellar
orbital motions. It is indeed quite instructive to see that only direct 
observations of the stellar absorbtion features of the M star enable
a proper interpretion of emission line radial velocities.

An even higher mass for the hot companion was suggested by Skopal et al. (2000), 
working with the same mass function as Thackeray \& Hutchings (1974). They assumed 
that the system
undergoes Roche lobe overflow. However, we have shown that the mass
function derived from emission lines is most likely seriously wrong, and that the 
Roche lobe assumption is very questionable.

\begin{table}
\caption[]{Roche lobe filling factors for symbiotic systems (see text).}
\label{roloff}
\begin{flushleft}
\begin{tabular}{lrcrcl}
\hline\noalign{\smallskip}
\ts\ts Object & Period & Separation & $R\tief c$~     & RL filling& Ref  \\
              & [d]~~   &  [AU]      & [\Rsolar]& factor \\
\hline\noalign{\smallskip}
\ts\ts AR Pav   & 604.5     & 1.95  & 130 & 0.52 & 8) \\  
\ts\ts FG Ser   & 650~~     & 1.95  & 105 & 0.41 & 7) \\
\ts\ts SY Mus   & 624.5     &  1.7  & 86  & 0.38 & 4) \\
\ts\ts EG And   & 482.2     &  1.5  & 75  & 0.37 & 3) \\
\ts\ts RW Hya   & 370.4     &  1.3  & 60  & 0.34 & 5) \\
\ts\ts AX Per   & 680.8     &  1.7  & 110 & 0.30 & 1) \\
\ts\ts AG Peg   & 816.5     &  2.5  & 85  & 0.25 & 2) \\
\ts\ts BX Mon   & 1401~~    & 2.0-5.9 & 160 & 0.68-0.18  & 6) \\
\hline\noalign{\smallskip}
\end{tabular}
\smallskip
1) Mikolajewska \& Kenyon (1992)\\ 
2) Kenyon et al.\ (1993)\\
3) Vogel et al.\ (1992)\\
4) Paper I\\ 
5) Paper II\\
6) Paper IV\\ 
7) Paper V\\
8) this paper 
\vskip-0.4cm
\end{flushleft}
\end{table}

One of the fundamental questions in relation to symbiotic binaries
is the process of mass transfer. Is it predominantly through
wind accretion or do the red giants fill their Roche lobes? 
We now have a small, but already significant sample of symbiotic systems
with accurate orbits, they allow to draw some preliminary conclusions. 
In Tab.~\ref{roloff}
we have summarised what we consider to be the presently best available 
values. The Roche lobe filling factor is the ratio of the red giant's
radius to the distance to the inner Lagrange point. We note that
none of the symbiotic red giants fills its Roche lobe. A similar
result was already obtained by M\"urset \& Schmid (1999), who analysed
a larger sample of symbiotic stars with, however, less accurately known
red giant radii and binary configurations. We therefore conclude that mass 
transfer occurs predominantly through wind accretion. 

Related to this accretion process is the question whether the accretor
is surrounded by an accretion disk. To model the wind accretion onto
the white dwarf it is necessary to know the physical conditions in its
neighbourhood. Among the crucial parameters are the white dwarf
radiation field as well as its magnetic field. A strong enough white
dwarf radiation field could prevent accretion, a
magnetic field could lead to accretion downflows over the polar
regions. For AR~Pav, unfortunately nothing is known about 
the white dwarf luminositiy and on its 
magnetic field. Wind accretion models which neglect both
effects such as those of Mastrodemos \& Morris~1998 and 
Mastrodemos \& Morris~1999 suggest the formation of accretion 
disks for a wide range of geometric configurations and 
red giant wind parameters. Some of their models are 
similar to those of AR~Pav. With the observed
radial velocity behaviour of the permitted emission lines,
it is likely that at least part of these emission lines originate
in the vicinity of the white dwarf. Applying eclipse techniques to
available and new high resolution emission line profiles with
good orbital coverage, to search
for spectral traces of an accretion disk and to study the 
circumstellar matter distribution would certainly be an interesting 
continuation to this work.

\begin{acknowledgements}
We are indebted to the ESO staff at Garching and La Silla who 
made the remote observations possible. TD acknowledges
financial support by the Swiss National Science Foundation, and HMS by
the Deutsche Forschungsgemeinschaft (WO 296/27-1).
\end{acknowledgements}

\end{document}